\documentclass[11pt,letter]{article}

\usepackage{graphicx}
\usepackage{amsmath}
\usepackage{amssymb}
\usepackage{cancel}
\usepackage[left=2cm,right=2cm,top=3cm,bottom=2cm]{geometry}
\usepackage{setspace}
\usepackage{float}
\usepackage[super,sort&compress,comma]{natbib}
\usepackage[font={footnotesize}]{caption}
\usepackage[labelfont=bf]{caption}
\usepackage{bm}
\usepackage[T1]{fontenc}
\usepackage{microtype}
\usepackage{subfigure}
\usepackage{chemformula}
\usepackage[version=4]{mhchem}
\usepackage{textcomp}

\usepackage{color}
\include{MyCommand}

\usepackage{tabularx}
\usepackage{booktabs}
\usepackage{multirow}
\usepackage{mathtools}
\usepackage{bm}
\usepackage{gensymb}
\usepackage{esvect}
\usepackage{makecell}

\begin{document}
\title{\textbf{ECENet: An Edge Cluster Expansion Line-Graph Neural Network}}
\date{}
\author{R. Allen LaCour*$^{1,2}$, Teresa Head-Gordon*$^{1,2,3}$}
\maketitle
\noindent
\begin{center}
$^1$Kenneth S. Pitzer Theory Center and Department of Chemistry\\
$^2$Chemical Sciences Division, Lawrence Berkeley National Laboratory\\
$^3$Departments of Bioengineering and Chemical and Biomolecular Engineering\\
University of California, Berkeley, CA, 94720 USA

corresponding authors: alacour@berkeley.edu, thg@berkeley.edu
\end{center}

\begin{abstract}
\noindent
Machine-learned interatomic potentials (MLIPs) have emerged as a promising alternative to classical force fields and first-principles theory for predicting the properties of chemical and material systems. Many MLIPs are graph neural networks with O(3)-equivariant features, whose accuracy comes at substantial computational cost. Here we introduce the edge cluster expansion (ECE), an analogue of the atomic cluster expansion in which the environment is expanded around edges between atom pairs rather than single atoms, and build upon it to develop the line-graph neural network ECENet. ECENet uses O(2)-equivariant features that persist on the edges between atoms, giving it natural access to O(2) operations that are cheaper and less restrictive than their O(3) counterparts. ECENet performs at the state of the art on the MD22 benchmark and lies on the accuracy–cost Pareto frontier when trained on the SPICE-MACE-OFF dataset. Furthermore, with long-range electrostatics implemented via latent Ewald summation, ECENet accurately predicts molecular dipole moments and the infrared spectrum of liquid water. The frontier ECENet architecture establishes equivariant edge-centered representations as an efficient and physically expressive foundation for equivariant MLIPs.
\end{abstract}

\newpage

\section{Introduction}
\vspace{-2mm}

Molecular dynamics (MD) is a central technique for studying the equilibrium and dynamic behavior of chemical and material systems. Key to its efficacy is an accurate model for interatomic energies and forces whose computational expense is low enough to allow for converged sampling. The most common approaches are to compute energies and forces with expensive \emph{ab initio} methods such as density functional theory (DFT)\cite{hohenberg1964inhomogeneous, kohn1965self, car1985unified, marx2009ab}, which limits MD to short simulations of small systems, or with inexpensive classical force fields (FFs)\cite{cornell1995second, jorgensen1996development, mackerell1998all, wang2004development, albaugh2016advanced}, whose fixed functional forms are less accurate but enable simulations over longer timescales and larger length scales.\cite{Demerdash2018}

Machine-learned interatomic potentials (MLIPs) have recently emerged\cite{behler2007generalized, bartok2010gaussian, deringer2019machine, unke2021machine} as an attractive alternative to both \emph{ab initio} and FF potential energy surfaces. Like FFs, MLIPs are parametrized functions of the atomic positions, but their functional forms are flexible enough to reproduce \emph{ab initio} energies and forces to near-chemical accuracy at a small fraction of the cost. Most MLIPs are graph neural networks in which each atom carries a feature vector that is updated by message passing with its neighbors\cite{gilmer2017neural, schutt2018schnet}.
Alternatively, an MLIP can be built on the line graph, in which the edges of the atomic graph carry the features and pass messages through the nodes\cite{gasteiger2021directional, choudhary2021atomistic, musaelian2023learning}.
The features may be invariant under rotations of the system\cite{smith2017ani, drautz2019atomic}, such as distances and angles, or equivariant under them\cite{batzner2022e3, batatia2022mace, schutt2018schnet, liao2023equiformer, haghighatlari2022newtonnet}, such as displacement vectors and higher-order spherical harmonics.
Equivariant models currently lead most MLIP accuracy benchmarks\cite{batzner2022e3, batatia2022mace, fu2025learning}, but they are usually considerably slower than their invariant counterparts.

Two factors constrain the utility of MLIPs in scientific applications. First, because MLIPs encode few physical priors, they require large amounts of training data and may behave unphysically outside it. The recent publication of increasingly large and diverse high-quality datasets\cite{Barroso-Luque_2026_omat24, levine2025omol25, Wood_2025_uma} has lessened this problem and has led to robust ``universal'' models\cite{chen2022universal, deng2023chgnet, batatia2025foundation, yuan2026foundation} that can simulate systems of broad elemental and configurational diversity. Nonetheless, discrepancies between the predictions of these models and experiment, such as liquid densities\cite{batatia2026macepolar}, point to the need both for more high-quality data and for more physically motivated model components, such as explicit long-range electrostatics, that generalize beyond the training input. 

The second constraining factor is their computational efficiency. Although far faster than \emph{ab initio} MD, MLIPs remain slow relative to classical force fields.
The slow simulation speed of MLIPs stems from the same expressive functional forms with many adjustable weights that make them accurate.
Furthermore, maintaining equivariance throughout the network permits only a limited set of operations, such as equivariant linear maps and gated nonlinearities, and the Clebsch--Gordan tensor product at the heart of most equivariant networks scales steeply with the degree of angular momentum coupling.\cite{passaro2023reducing, batzner2022e3} Hence, a current frontier in MLIP development is to formulate architectures that retain equivariance at lower computational cost. One widely adopted approach, introduced in eSCN\cite{passaro2023reducing} and since used in various forms by other architectures\cite{liao2024equiformerv2, fu2025learning, Wood_2025_uma, li2026dpa4, xu2026edge}, exploits the SO(2) symmetry about each graph edge by rotating features into an edge-aligned frame, where the SO(3) tensor product reduces to cheaper SO(2) operations. Even so, these models exploit SO(2) operations only transiently during message passing, and otherwise maintain SO(3)-equivariant atomic features with the associated expense.

Here we introduce ECENet, an equivariant line-graph neural network built on the edge cluster expansion (ECE). Instead of maintaining atomic features, ECENet maintains persistent features only in the edge frame, which are O(2)-equivariant by construction.
The total number of features therefore scales with the number of edges rather than the number of atoms, increasing expressivity, while computational expense is limited by the use of efficient and expressive O(2) operations throughout the network.
In what follows, we first discuss the ECE formalism and then its incorporation into the ECENet line-graph architecture. We show that ECENet trained from scratch matches the best current model on the MD22 benchmark, and that ECENet models trained on the SPICE-MACE-OFF dataset lie on the accuracy--efficiency Pareto frontier alongside the strongest current model on that dataset, DPA4\cite{li2026dpa4}. Next, we show that including long-range electrostatics via the latent Ewald summation (LES) approach, as introduced by Cheng\cite{cheng2025latent}, enables ECENet to predict molecular dipole moments and the infrared spectrum of liquid water without ever being trained on electrostatic multipoles. These results establish equivariant line-graph networks as a physically expressive route to MLIPs with frontier-level accuracy and efficiency.

\section{Model}
\subsection{The edge cluster expansion and ECENet}
\vspace{-2mm}

ECE's core idea of exploiting O(2) symmetry in the edge frame was introduced in passing by Dusson et al. \cite{dusson2022atomic} under the name ``bond cluster expansion'' in their study of the atomic cluster expansion (ACE)\cite{drautz2019atomic}.
Later, Zitnick and co-workers\cite{passaro2023reducing} showed that the SO(2) symmetry of an edge-aligned frame can be used to accelerate the Clebsch--Gordan tensor product at the heart of most equivariant message-passing MLIPs, and subsequent work found further uses for operations unique to SO(2)\cite{fu2025learning, li2026dpa4}, with Xu et al.\cite{xu2026edge} describing their construction as an ``edge cluster expansion'' (ECE). We also develop our model within the ECE framework but take a different approach: rather than using the edge frame only to evaluate messages between atoms, we make O(2)-equivariant edge features the primary and persistent representation of the network.

\begin{figure}[H]
\centering
\includegraphics[width=0.85\textwidth]{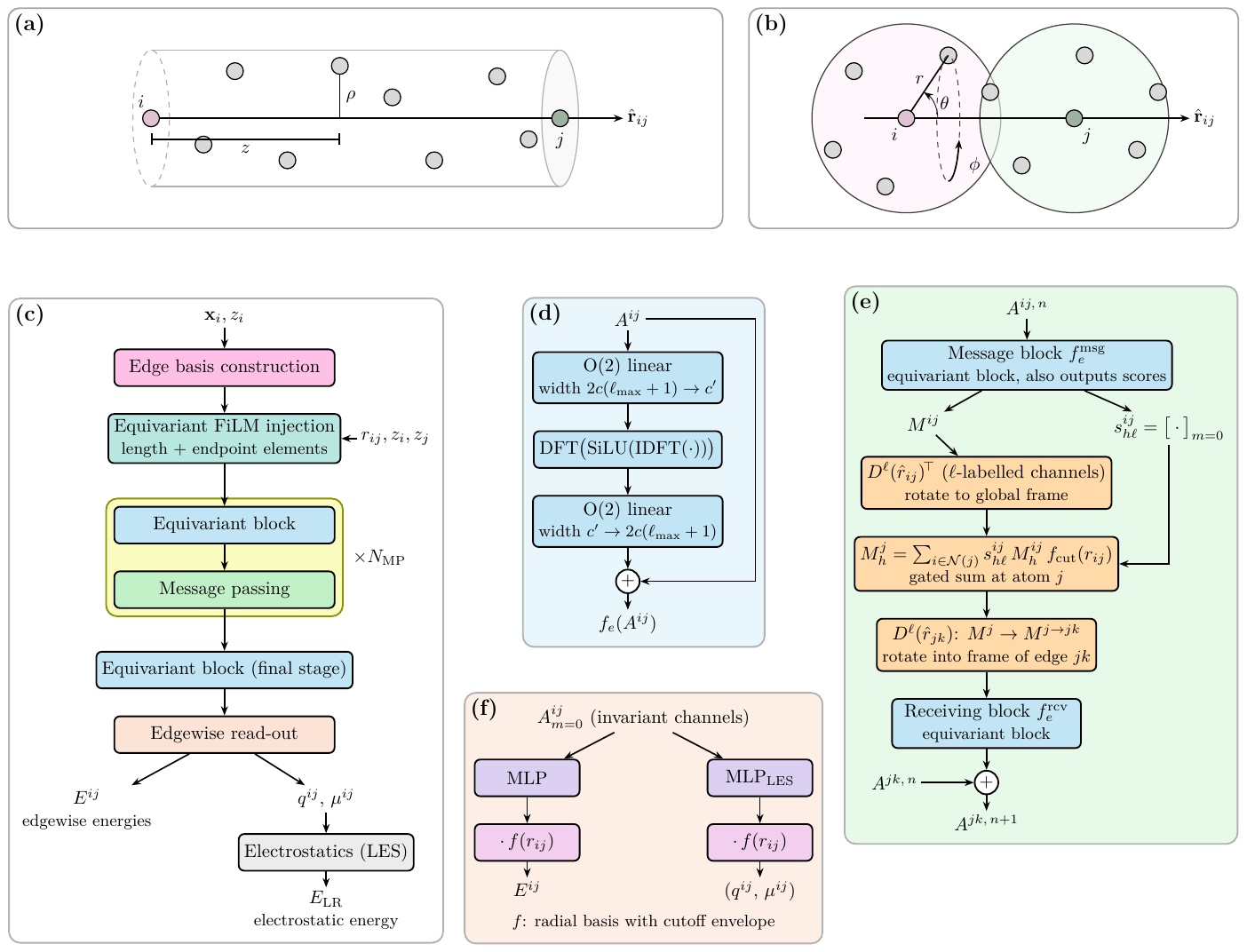}
\vspace{-2mm}
\caption{\textbf{The ECENet equivariant line-graph model with O(2) edge embeddings}. The ECE basis is built from circular harmonics using either (a) the cylindrical pair or (b) a concatenation of atom-centered bases. (c) Schematic of the entire ECENet architecture. (d) The equivariant block, (e) the message-passing block, and (f) the edgewise readout with LES\cite{cheng2025latent} for long-range electrostatics.}
\label{fig:architecture}
\end{figure}

ECE is analogous to ACE\cite{drautz2019atomic} in that the environment is expanded in a body-ordered basis but is distinguished by the expansion being centered on an edge between a pair of atoms rather than on a single atom. In ACE, the basis is built from spherical harmonics so that it transforms according to irreducible representations of O(3). Fixing an edge direction reduces this symmetry to the subgroup that leaves the axis invariant, O(2), which encompasses rotations about the edge axis and reflections in planes containing it. The ECE basis is accordingly built from circular harmonics. 
Because circular harmonics account for only one angular degree of freedom, two further coordinates are needed to specify a neighbor's position, compared with the single radial coordinate of ACE. One natural choice is the cylindrical pair: the signed distance along the edge, z, and the distance from the edge axis, $\rho$ (Figure \ref{fig:architecture}a). Alternatively, an edge basis can be formed by concatenating atom-centered bases of the ACE type evaluated at each endpoint atom (Figure \ref{fig:architecture}b), as described below.

The total number of features in ECE scales as $Nd$, where $N$ is the number of atoms and $d$ is the average number of neighbors per atom.
This scaling is greater than ACE's $N$, and thus ECE may be more expressive than ACE at the same number of parameters. On the other hand, each edge basis depends on the $\sim d$ atoms in its environment, and therefore naively computing the ECE features requires $\mathcal{O}(Nd^2)$ basis-function evaluations, compared with $\mathcal{O}(Nd)$ in ACE.
However, using a concatenation of ACE-type basis functions reduces the complexity. 
When the polar axis of the spherical harmonics is aligned with the edge, the azimuthal factor of $Y_{\ell m}$ reduces to a circular harmonic of frequency $m$, independent of $\ell$.
This alignment can be performed exactly by rotation with Wigner $D$-matrices; one can thus compute $N$ atomic bases in a global frame, as in ACE, and obtain the $Nd$ edge bases by rotating each atomic base into the frame of each of its $d$ edges.
Since applying a Wigner $D$-matrix is inexpensive relative to evaluating basis functions, the cost of computing ECE's basis functions is on the order of $Nd$ rather than $Nd^{2}$.

In linear ACE, an MLIP is constructed from symmetrized products of the atomic base, contracted into invariants. An analogous linear ECE model can be built from the ECE basis in a manner that highlights the benefit of the $O(2)$ structure of its features. When the tensor product is taken between two irreps of $O(2)$ of frequencies $a$ and $b$, the result decomposes into only two irreps, of frequencies $|a-b|$ and $a+b$, and the mixing is trivial (diagonal in the complex basis, since $e^{ia\varphi}e^{ib\varphi} = e^{i(a+b)\varphi}$). In contrast, the tensor product of $O(3)$ irreps of degrees $\ell=a$ and $\ell=b$ decomposes into irreps of every degree from $|a-b|$ to $a+b$, each appearing exactly once, with mixing given by Clebsch--Gordan coefficients.
Symmetrized products in ECE therefore reduce to addition and subtraction of frequencies, with no Clebsch--Gordan coupling at all.


Rather than being linear models, most recent MLIPs, including MACE\cite{batatia2022mace}, DPA4\cite{li2026dpa4}, and UMA\cite{Wood_2025_uma}, incorporate an ACE-like atomic basis into a graph neural network. Message passing between atoms propagates information beyond the local cutoff and composes features nonlinearly across layers. As mentioned previously, models of this type currently lead most benchmarks. With ECENet, we instead incorporate the O(2)-equivariant edge features of ECE into a line-graph neural network, in which the edges of the atomic graph carry the features and pass messages through the atoms. As described below, this allows the O(3)-equivariant operations of atom-centered networks to be replaced by O(2)-equivariant analogues that are cheaper and, in several respects, more expressive. The overall model is illustrated in Figure \ref{fig:architecture}c. Below, we deconstruct the operations of ECENet in more detail.

\subsection{Edge Construction}
\vspace{-2mm}

To construct the initial edge features, we first build an atomic base $A^{i}$ for each atom, as in ACE. The one-particle basis functions are products of radial functions $R_n$, here sinc functions multiplied by a smooth cutoff envelope, and real spherical harmonics up to degree $\ell_{\max}$, evaluated in a global frame:
\begin{equation}
    A^{i}_{en\ell m} = \sum_{\substack{j \in \mathcal{N}(i) \\ e_j = e}} R_n(r_{ij})\, Y_{\ell m}(\hat{\mathbf{r}}_{ij}),
    \label{eq:atomic_base}
\end{equation}
where $n$ indexes the radial basis, $e$ the element of the neighbor, and $\mathcal{N}(i)$ the neighbors of atom $i$ within the cutoff radius $r_c$. We then contract the radial and element indices with learned weights, specific to the element $z_i$ of the central atom, to form $c$ channels per $(\ell, m)$:
\begin{equation}
    A^{i}_{c\ell m} = W_{z_i,\,cen}\, A^{i}_{en\ell m}.
    \label{eq:embed}
\end{equation}
We form edge features for each ordered pair of atoms $(i,j)$ within the same cutoff radius $r_c$ by rotating both atomic bases into a common frame whose polar axis is aligned with the edge and then concatenating them:
\begin{equation}
    A^{ij}_{c\ell m} = \left[ D^{\ell}(\hat{\mathbf{r}}_{ij})\, A^{i}_{c\ell m},\; D^{\ell}(\hat{\mathbf{r}}_{ij})\, A^{j}_{c\ell m} \right],
    \label{eq:edge_basis}
\end{equation}
where $D^{\ell}(\hat{\mathbf{r}}_{ij})$ denotes the Wigner $D$-matrix of the rotation taking the global $z$-axis to the edge direction $\hat{\mathbf{r}}_{ij}$. 
We use directed edges because they handle the exchange of endpoints naturally, and thus each pair of atoms yields two directed edges, $A^{ij}$ and $A^{ji}$ (the latter using $\hat{\mathbf{r}}_{ji} = -\hat{\mathbf{r}}_{ij}$). 
In the edge frame, the residual symmetry is O(2), which corresponds to all rotations about the edge axis and reflections through planes containing it. 
Under O(2), features transform only according to their azimuthal frequency $m$, with the degree $\ell$ acting as an ordinary channel index rather than a symmetry label.
We retain frequencies $|m| \le m_{\max}$, where $m_{\max} \le \ell_{\max}$.
The values of $m_{\max}$ and $\ell_{\max}$ are both hyperparameters important for the cost and accuracy of the network. 
We further exploit the demotion of $\ell$ by rectangularizing the feature grid: each frequency block, which natively contains $2c(\ell_{\max} - |m| + 1)$ components, is padded to a uniform width of $2c(\ell_{\max}+1)$ channels, initialized to zero and populated by the first equivariant block. 
Because the added channels carry the same frequency $m$ as their block, equivariance is preserved, and the uniform block size allows all frequencies to be processed with identical batched operations.

After forming the edge basis, we inject edge length and endpoint element information into $A^{ij}$ using feature-wise linear modulation (FiLM)~\cite{perez2018film}.
Each channel is multiplied by a learned scaling factor, with identical factors applied to the $+m$ and $-m$ channels, and the $m=0$ channels additionally receive a learned shift; because scaling commutes with rotations about the edge axis and shifts affect only the invariant frequency, this modulation preserves $O(2)$ equivariance.
The scaling and shift factors are produced by a small multilayer perceptron applied to the concatenation of learned element embeddings for atoms $i$ and $j$ and a radial-basis expansion of the edge length $r_{ij}$.
Since the edges are directed, the ordered concatenation of the two element embeddings makes the modulation sensitive to edge orientation.
Although the atomic bases already encode radial and element information, they aggregate over all neighbors of each endpoint and therefore contain nothing that singles out the edge itself; the FiLM layer supplies this missing information, conditioning the edge features on their own length and endpoint elements.

\subsection{Equivariant Block}
\vspace{-2mm}

After the FiLM injection, we apply an $O(2)$-equivariant block (Figure \ref{fig:architecture}d), consisting of a nonlinearity $\tilde{\sigma}$ between two $O(2)$-equivariant linear layers with a skip connection:
\begin{equation}
f_e(A^{ij}) \;=\; A^{ij} + W_2^{O(2)}\,\tilde{\sigma}\!\left(W_1^{O(2)} A^{ij}\right),
\label{eq:equivariant_block}
\end{equation}
where $W_1^{O(2)}$ expands (or contracts) the channel dimension, $\tilde{\sigma}$ acts at the altered width, and $W_2^{O(2)}$ restores the original width so that the skip connection is well defined.
An O(2)-equivariant linear layer can mix freely across channels and across $\ell$ at fixed $|m|$, provided the $+m$ and $-m$ components receive the same real weight block (a complex weight would be SO(2)-equivariant but not invariant under the reflections of O(2)), and it may carry a bias on the invariant $m = 0$ components only. 
By contrast, an $O(3)$-equivariant linear layer cannot mix across $\ell$, and all $2\ell+1$ components of $m$ at a given $\ell$ must share a single weight, making it considerably more restrictive than its $O(2)$ counterpart.

For the nonlinearity $\tilde{\sigma}$, we use a Fourier pointwise activation as an alternative to explicit tensor products. The frequency-space features are transformed to values on a uniform grid of $N$ points over the azimuthal angle, a pointwise nonlinearity is applied on the grid, and the result is transformed back:
\begin{equation}
    \tilde{\sigma}(A^{ij}) = \mathrm{DFT}\!\left( \mathrm{SiLU}\!\left( \mathrm{IDFT}(A^{ij}) \right) \right),
    \label{eq:fourier_activation}
\end{equation}
where the transforms act over the azimuthal angle independently for each channel.
A pointwise
activation commutes with both a cyclic shift and a reversal of the grid, so
the output transforms exactly as the input does. The equivariance is exact
up to aliasing;
here, we always use $m_{\max} = 2$ with at least 10 grid points, twice the minimum of $2m_{\max}+1$ points needed to represent the input. 
On the grid, pointwise powers of the features correspond to repeated convolutions in frequency space, so the activation implicitly forms products of many orders, filling the role played by explicit tensor products in linear ECE, at the cost of two linear transforms and a pointwise operation. Analogous grid activations are used in O(3)-equivariant models~\cite{cohen2018spherical,zitnick2022spherical,passaro2023reducing,liao2024equiformerv2}, but they require quadrature over the sphere, whereas in the edge frame a one-dimensional transform over the circle suffices.

\subsection{Message Passing}
\vspace{-2mm}

Messages are passed edge-to-edge through shared endpoint atoms: each edge emits a message, incoming messages are aggregated at each atom, and the aggregate is redistributed to the edges leaving that atom (Figure \ref{fig:architecture}e). 
This is message passing on the line graph of the atomic graph, whose nodes are the directed atomic edges and whose edges connect atomic edges that share an atom; the atoms themselves serve only as aggregation sites.
Rather than conditioning messages on an explicit function of bond length and endpoint elements, we produce each message by applying a dedicated equivariant block of the form of equation \ref{eq:equivariant_block} to the edge features:
\begin{equation}
    M^{ij} = f_e^{\mathrm{msg}}(A^{ij}),
    \label{eq:message}
\end{equation}
so that $M^{ij}$ has the same shape as $A^{ij}$. The output layer of this block additionally emits $H(\ell_{\max}+1)$ channels beyond the feature width, and the invariant ($m = 0$) component of each provides a scalar score $s^{ij}_{h\ell}$ for gating head $h$ and degree $\ell$.
Using the Wigner $D$-matrices computed during edge construction, each message is rotated from its edge frame back into the global frame, where messages from different edges transform consistently and can be summed equivariantly. Because transport between frames acts through the $D^{\ell}$-matrices, only components carrying a definite degree $\ell$ can be rotated: the $\ell$-labeled (triangular) subset of $M^{ij}$ is transported, while the padded channels remain in the edge frame and participate only through the per-edge blocks. The transported message is subdivided along the channel axis into $H$ blocks $M^{ij}_{h}$, one per head, and the aggregate at atom $j$ is
\begin{equation}
    M^{j}_{h,\ell m} = \sum_{i \in \mathcal{N}(j)} s^{ij}_{h\ell}\, M^{ij}_{h,\ell m}\, f_{\mathrm{cut}}(r_{ij}),
    \label{eq:aggregate}
\end{equation}
where $f_{\mathrm{cut}}$ is the same smooth cutoff function used in the radial basis, ensuring continuity as atoms enter and leave the neighborhood. The aggregate is then distributed to each edge $jk$ leaving atom $j$: it is rotated into that edge's frame, denoted $M^{j \to jk}$, passed through a receiving equivariant block of the form of equation \ref{eq:equivariant_block}, and added residually to the edge features:
\begin{equation}
    A^{jk,\,n+1} = A^{jk,\,n} + f_e^{\mathrm{rcv}}\!\left( M^{j \to jk} \right),
    \label{eq:receive}
\end{equation}
where $n$ indexes the message-passing layer. 
No exclusion is applied to the reversed edge: the aggregate at atom $j$ includes the message from edge $kj$ and is distributed to edge $jk$, so each directed edge also receives, through the shared atom, information from its own reverse. The full network consists of stages of per-edge equivariant blocks separated by message-passing layers. Our use of O(3)-structured features in the global frame during aggregation on the atoms and O(2)-structured features in the edge frame parallels recent models that reduce SO(3) convolutions to SO(2) operations in an edge-aligned frame~\cite{passaro2023reducing,liao2024equiformerv2,fu2025learning,Wood_2025_uma,li2026dpa4}. Our approach is distinguished by the persistence of the edge features across layers: atoms serve only as transient aggregation sites, and no atomic features are maintained between layers.

\subsection{Readout}
\vspace{-2mm}

The total energy is a sum of per-atom and per-edge contributions. The per-atom terms $E_0(z_i)$, which depend only on the element $z_i$, are computed by linear regression of total configuration energies against element counts and fine-tuned during training. The per-edge contribution is computed from the edge features after the final stage: the invariant ($m = 0$) channels are passed through an MLP, and the result is contracted with a vector of radial basis functions of the edge length:
\begin{equation}
    E_{ij} = \mathrm{MLP}\!\left( A^{ij}_{m=0} \right) \cdot \mathbf{f}(r_{ij}).
    \label{eq:readout}
\end{equation}
As before, the radial basis carries the smooth cutoff envelope, so that $E_{ij}$ and its first derivative vanish continuously as $r_{ij}$ approaches the cutoff.
The total non-LES energy is
\begin{equation}
    E_{\mathrm{SR}} = \sum_{(i,j)} E_{ij} + \sum_i E_0(z_i),
    \label{eq:total_energy}
\end{equation}
where the first sum runs over directed edges. Forces are obtained as the negative gradient of the total energy with respect to atomic positions, computed by automatic differentiation, and are therefore conservative by construction.

\subsection{Latent Charges and Dipoles}
\vspace{-2mm}

We incorporate long-range electrostatics using the LES approach~\cite{cheng2025latent,King_2025_charges}, as shown in Figure \ref{fig:architecture}f. 
We use an expression analogous to equation~\ref{eq:readout}:.
\begin{equation}
    \left( q_{ij},\, \mu_{ij} \right) = \mathrm{MLP}_{\mathrm{LES}}\!\left( A^{ij}_{m=0} \right) \cdot \mathbf{f}(r_{ij}),
    \label{eq:les_edge}
\end{equation}
where the MLP output supplies one coefficient vector per predicted quantity, and the cutoff envelope in $\mathbf{f}$ ensures both quantities vanish smoothly at the cutoff. LES operates on per-atom charges and dipoles, so we aggregate over incoming edges,
\begin{equation}
    q_j = \sum_{i \in \mathcal{N}(j)} q_{ij}, \qquad
    \boldsymbol{\mu}_j = \sum_{i \in \mathcal{N}(j)} \mu_{ij}\, \hat{\mathbf{r}}_{ij},
    \label{eq:les_atom}
\end{equation}
where $\hat{\mathbf{r}}_{ij}$ is the unit vector pointing from atom $i$ to atom $j$. In the case of the dipole, multiplication by $\hat{\mathbf{r}}_{ij}$ promotes the scalar $\mu_{ij}$ to a vector with the correct rotational symmetry. The long-range energy $E_{LR}$ is then evaluated from the latent charges and dipoles using the smeared charge--charge, charge--dipole, and dipole--dipole interactions of references \citenum{cheng2025latent} and \citenum{kim2026polarizable}, and the total energy is $E = E_{SR} + E_{LR}$.

\section{Results}
\subsection{Interaction Range}
\vspace{-2mm}

\begin{figure}[H]
\centering
\includegraphics[width=0.8\textwidth]{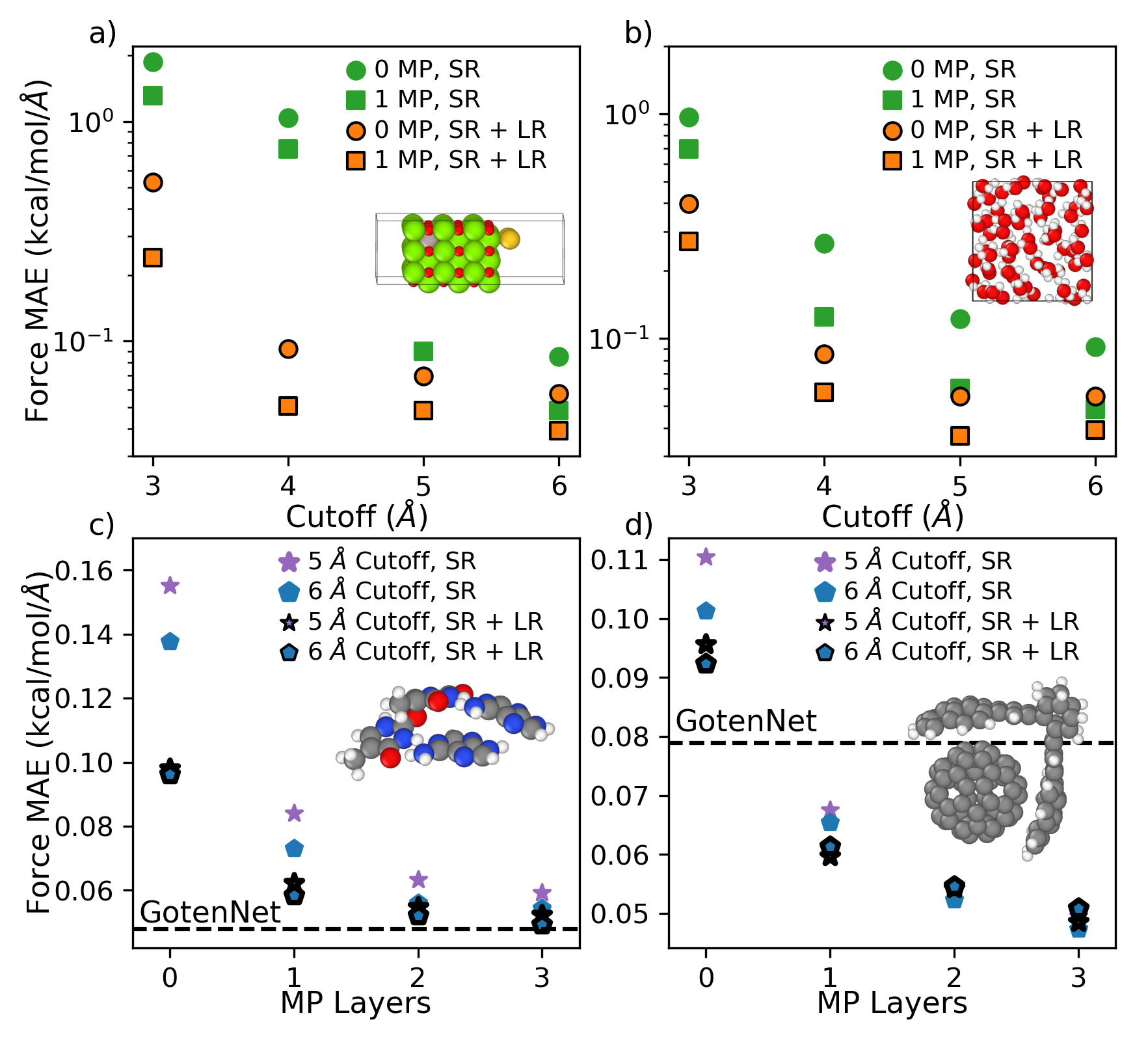}
\vspace{-2mm}
\caption{\textbf{Examining ECENet performance with interaction range}.
We examine the performance on  (a) Au$_2$ on MgO, (b) liquid water, (c) an AT-AT molecular complex, and (d) a buckyball catcher.
For the Au$_2$ on MgO and water systems, we examine how the force MAE varies with the cutoff, the presence of MP layers, and the addition of long-range interactions with LES.
For the AT-AT and buckyball catcher systems, we examine the same factors but limit our cutoffs to 5 \AA{} and 6 \AA{} while exploring more MP layers.}
\label{fig:range}
\end{figure}

Before applying ECENet to benchmark datasets, we first examined how the model captures interactions beyond the range of a single edge. Specifically, we varied three hyperparameters that control the interaction range: the cutoff radius, which determines both the edges and the neighborhoods entering the atomic bases; the number of message-passing (MP) layers; and the presence of long-range electrostatics via LES. We were particularly interested in message passing, since there is little precedent for equivariant message passing from edge to edge: Allegro~\cite{musaelian2023learning} deliberately omits it, and the edge-to-edge messages of other line-graph models are invariant. We examined four systems in which long-range effects are expected to be relevant for various reasons: liquid water~\cite{schmiedmayer2024derivative}, Au$_2$ on Al-doped MgO~\cite{ko2021fourth}, and the AT-AT and buckyball catcher complexes from the MD22 benchmark~\cite{chmiela2023accurate}. Figure~\ref{fig:range} reports the force MAE for each system as a function of these choices.

\vspace{-2mm}

Our results for Au$_2$ on MgO and liquid water are shown in Figures~\ref{fig:range}a and~\ref{fig:range}b, respectively.
Long-range electrostatics are important in both systems: in water because of its large molecular dipole, and in Au$_2$/MgO because adsorption behavior is governed by long-range charge transfer between the gold dimer and the Al-doped substrate.
Consistent with this expectation, at short cutoffs (3--4~\AA) adding a MP layer and adding LES both lower the force MAE, with LES having the larger effect.
Increasing the cutoff from 3~\AA{} to 4~\AA{} also sharply reduces the error, while further increases to 5~\AA{} or 6~\AA{} yield diminishing returns.
At the longest cutoff examined (6~\AA), the additional benefit of MP layers or LES is less than at shorter cutoffs.
This saturation is expected because the simulation cells are only $\sim$12~\AA{} across, so a 6~\AA{} cutoff already places most of the cell within each edge's receptive field, leaving little for message passing or LES to contribute.

Our results for the AT-AT and buckyball catcher systems are shown in Figures~\ref{fig:range}c and ~\ref{fig:range}d, respectively.
These systems are large isolated molecules whose spatial extent exceeds the cutoff, so interactions exist between atom pairs that no single edge environment contains, and dispersion is expected to be the dominant long-range interaction (especially for the buckyball catcher). To probe ECENet's ability to capture these longer-range interactions, we examined only cutoffs of 5~\AA{} and 6~\AA{} and vary the number of MP layers and the presence of LES.
For 0 or 1 MP layers, LES substantially lowers the force MAE, but its effect is weaker than adding an MP layer, and for the buckyball catcher---composed only of C and H, with weak electrostatics---LES becomes neutral or slightly detrimental once two or more MP layers are present.
The force MAE does consistently drop when going from a cutoff of 5 \AA{} to 6 \AA{}, but again the drop is smaller than that from adding an MP layer.
The benefit of additional MP layers is not completely saturated by 3 layers, but the gain for each additional layer is reduced, indicating sufficient range to capture most interatomic interactions.
For reference, we indicate the force MAE of GotenNet, the previous top performer on these MD22 systems. ECENet nearly matches GotenNet on AT-AT and substantially surpasses it on the buckyball catcher.

\subsection{Comparative Performance on the MD22 Dataset}
\vspace{-2mm}

We next trained ECENet from scratch on each system in the MD22 dataset~\cite{chmiela2023accurate}, which consists of molecules and molecular complexes of 42--370 atoms, and compared our results to other MLIP architectures trained previously on the same benchmark. Each ECENet model used two MP layers, a 6~\AA{} cutoff, and
1.1M parameters.
We used the standard training-set sizes found in 
reference ~\citenum{chmiela2023accurate}; complete training details are given in the Methods. Table~\ref{tab:md22} compares energy and force MAEs against other top-performing models.

\begin{table}[H]
\centering
\setlength{\tabcolsep}{4pt}
\renewcommand{\arraystretch}{1.1}
\caption{\textbf{Comparative performance on the MD22 dataset among different MLIP architectures.} The units of energy (E) and forces (F) are kcal/mol and kcal/mol/Å, respectively, and E is per atom. The best results for each category is shown in bold. GotenNet$_B$\cite{aykent2025gotennet} and ECENet are the best-performing MLIPs, although the ECENet model contains far fewer parameters.}
\vspace{-2mm}
\resizebox{\textwidth}{!}{%
\begin{tabular}{|l|c|cccccccc|cc|}
\hline
Model & & Allegro\cite{musaelian2023learning, li2024longshort} & MACE\cite{kovacs2023evaluation} & Equiformer\cite{liao2023equiformer, li2024longshort} & ViSNet\cite{wang2024enhancing, li2024longshort} & QuinNet\cite{wang2023efficiently} & E-LSRM\cite{li2024longshort} & V-LSRM\cite{li2024longshort} & ESCAIP\cite{qu2024importance} & GotenNet$_B$\cite{aykent2025gotennet} & ECENet \\
\hline
Parameter Count & & 15M & 3.8M$^*$ & 3.0M & 2.2M & 8.8M$^*$ & 3.0M & 1.7M  & 15.0M & >9.2M$^*$ & 1.1M \\
\hline
\multirow{2}{*}{Tetrapeptide (42)} & 
 E &  0.1019 & 0.0620 & 0.0828 & 0.0796 & 0.0840 & 0.0780 & 0.0654 & 0.0589 & 0.0505 & \textbf{0.0438} \\
& F  & 0.1068 & 0.0876 & 0.0804 & 0.0972 & 0.0681 & 0.0887 & 0.0902 & 0.0719 & 0.0567 & \textbf{0.0514} \\
\hline
\multirow{2}{*}{DHA (56)} 
 & E & 0.1153 & 0.1317 & 0.1788 & 0.1526 & 0.1200 & 0.0878 & 0.0873 & 0.0640 & 0.0575 & \textbf{0.0497 }\\
 & F  & 0.0732 & 0.0646 & 0.0506 & 0.0668 & 0.0515 & 0.0534 & 0.0598 & 0.0496 & 0.0421 & \textbf{0.0333} \\
\hline
\multirow{2}{*}{Stachyose (87)} 
 & E  & 0.2485 & 0.1244 & 0.1404 & 0.1283 & 0.2300 & 0.1252 & 0.1055  & 0.0751 & 0.0673 & \textbf{0.0621} \\
 & F  & 0.0971 & 0.0876 & 0.0635 & 0.0869 & 0.0543 & 0.0632 & 0.0767 & 0.0512 & \textbf{0.0427} & 0.0441 \\
\hline
\multirow{2}{*}{AT-AT (60)} 
 & E  & 0.1428 & 0.1093 & 0.1309 & 0.1688 & 0.1400 & 0.1007 & 0.0772 & 0.0640 & \textbf{0.0544} & 0.0600 \\
 & F  & 0.0952 & 0.0992 & 0.0960 & 0.1070 & 0.0687 & 0.0881 & 0.0781  & 0.0632 & \textbf{0.0478} & 0.0529 \\
\hline
\multirow{2}{*}{AT-AT-CG-CG (118)} 
 & E & 0.3933 & 0.1578 & 0.1510 & 0.1995 & 0.3800 & 0.1335 & 0.1135 & 0.0964 & 0.0923 & \textbf{0.0908} \\
 & F & 0.1280 & 0.1153 & 0.1252 & 0.1563 & 0.1273 & 0.1065 & 0.1063 & 0.0824 & 0.0744 & \textbf{0.0485} \\
\hline
\multirow{2}{*}{\makecell[l]{Buckyball\\catcher (148)}}
 & E  & 0.5258 & 0.481 & 0.3978 & 0.4421 & 0.5624 & -- & 0.4220  & 0.3432 & 0.3032 & \textbf{0.2982} \\
 & F & 0.0887 & 0.085 & 0.1114 & 0.1335 & 0.1091 & -- & 0.1026  & 0.0838 & 0.0789 & \textbf{0.0587} \\
\hline
\multirow{2}{*}{\makecell[l]{Double-walled\\nanotube (370)}} 
 & E & 2.2097 & 1.655 & 1.1945 & 1.0339 & 1.8130 & -- & 1.8230  & 0.9993 & \textbf{0.6641} & 0.8491 \\
 & F & 0.3428 & 0.277 & 0.2747 & 0.3959 & 0.2473 & -- & 0.3391 & 0.2464 & \textbf{0.1888} & 0.2104 \\
\hline
\end{tabular}}
\par\vspace{1mm}
\begin{minipage}{\textwidth}
\footnotesize $^*$Parameter counts not reported by the original authors; instead estimated from reported hyperparameters. The reported 9.2M for GotenNet is a lower bound, as the models trained for MD22 were larger than those trained for QM9.
\end{minipage}
\label{tab:md22}
\end{table}
\vspace{-2mm}

GotenNet and ECENet emerge as the two best-performing models, splitting the top position across the seven systems. No obvious system property predicts which model wins: ECENet performs best on the second-largest system (the buckyball catcher), while GotenNet performs best on the largest system, the double-walled nanotube. Notably, the ECENet models are smaller than the other reported architectures, including GotenNet (1.1M vs. $>$9.2M parameters). We attribute this parameter efficiency to the edge-based representation: with $2d$ directed edges per atom, ECENet maintains roughly $2d$ times as many feature vectors as a node-based model of equal width. The parameters acting on them are shared across all edges, so representational state grows without a corresponding growth in parameters.

\subsection{Comparative Performance on the SPICE-MACE-OFF Dataset}
\vspace{-2mm}

We next evaluated ECENet on the SPICE-MACE-OFF dataset~\cite{Eastman_2023_spice,
kovacs2025}, the subset of SPICE used to train the original MACE-OFF models. It contains molecules and molecular complexes (dimers,
solvated amino acids, and water clusters) spanning ten elements (H, C, N, O,
F, P, S, Cl, Br, I), and models trained on it transfer to organic molecules
outside the training set~\cite{kovacs2025}, making it a standard
benchmark for general-purpose organic MLIPs.
We trained a smaller model,
ECENet-1.7M, with one MP layer, and a larger model, ECENet-4.8M, with three
MP layers (see Methods), and evaluated both on the standard per-category test set.

\begin{figure}[H]
\centering
\includegraphics[width=0.8\textwidth]{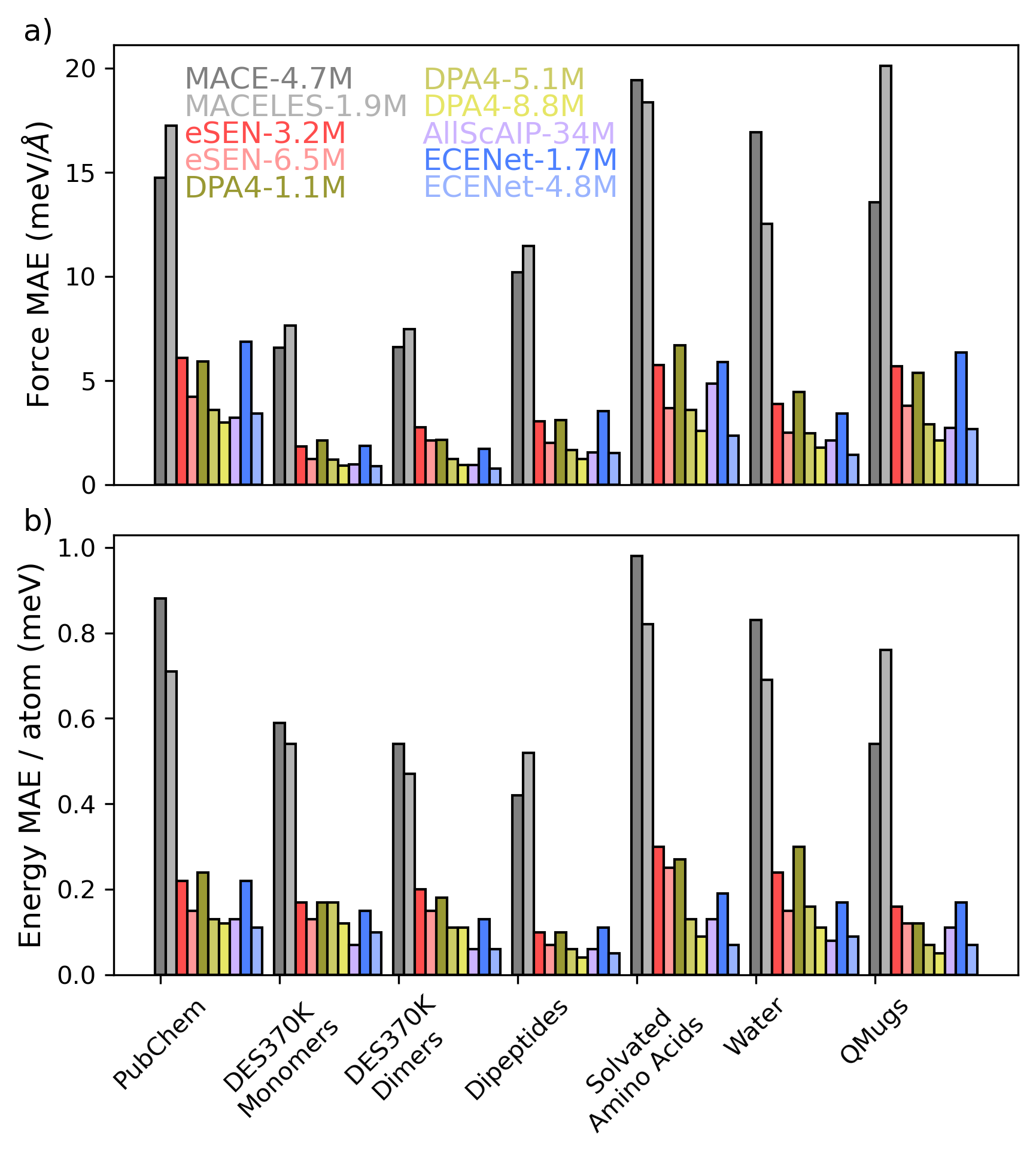}
\vspace{-3mm}
\caption{\textbf{Performance on the standard SPICE-MACE-OFF test set.} Comparison among current state-of-the-art MLIPs for (a) the force MAE and (b) the energy MAE per atom; see Supplementary Table S1 for the numerical values. On most splits, the larger ECENet model performs similarly to the current top performer on the dataset, DPA4, while the smaller ECENet model performs similarly to the other smaller models.}
\label{fig:spice_results}
\end{figure}

Figure~\ref{fig:spice_results} reports their energy and force MAEs alongside
models trained on the same data; see Supplementary Table S2 for tabular results. 
The other models shown are MACE-4.7M~\cite{kovacs2025}, from the original
model family; MACELES-1.9M\cite{kim2025universal}, which augmented MACE with LES;
and eSEN~\cite{fu2025learning}, DPA4~\cite{li2026dpa4} and
AllScAIP~\cite{qu2026recipe}, which represent the current state of the art.
We denote each variant by its number of parameters.
Unsurprisingly, the recent models all outperform the MACE-4.7M baseline.
ECENet-4.8M obtains excellent results on the dataset, having the lowest force MAE in four of seven categories and the lowest energy MAE in five of seven categories. 
Relative to DPA4-8.8M, previously the most accurate model, ECENet-4.8M is essentially tied in geometric-mean force MAE (1.65 vs. 1.64 meV/Å) and lower in energy (0.076 vs. 0.085 meV/atom), with 55$\%$ of the parameters. It has a lower force MAE than eSEN-6.5M in every category and than AllScAIP-34M in six of seven. 
At the small end, ECENet-1.7M has lower geometric-mean energy and force errors than both eSEN-3.2M and DPA4-1.1M,
the models closest to it in size.

\begin{figure}[H]
\centering
\includegraphics[width=0.8\textwidth]{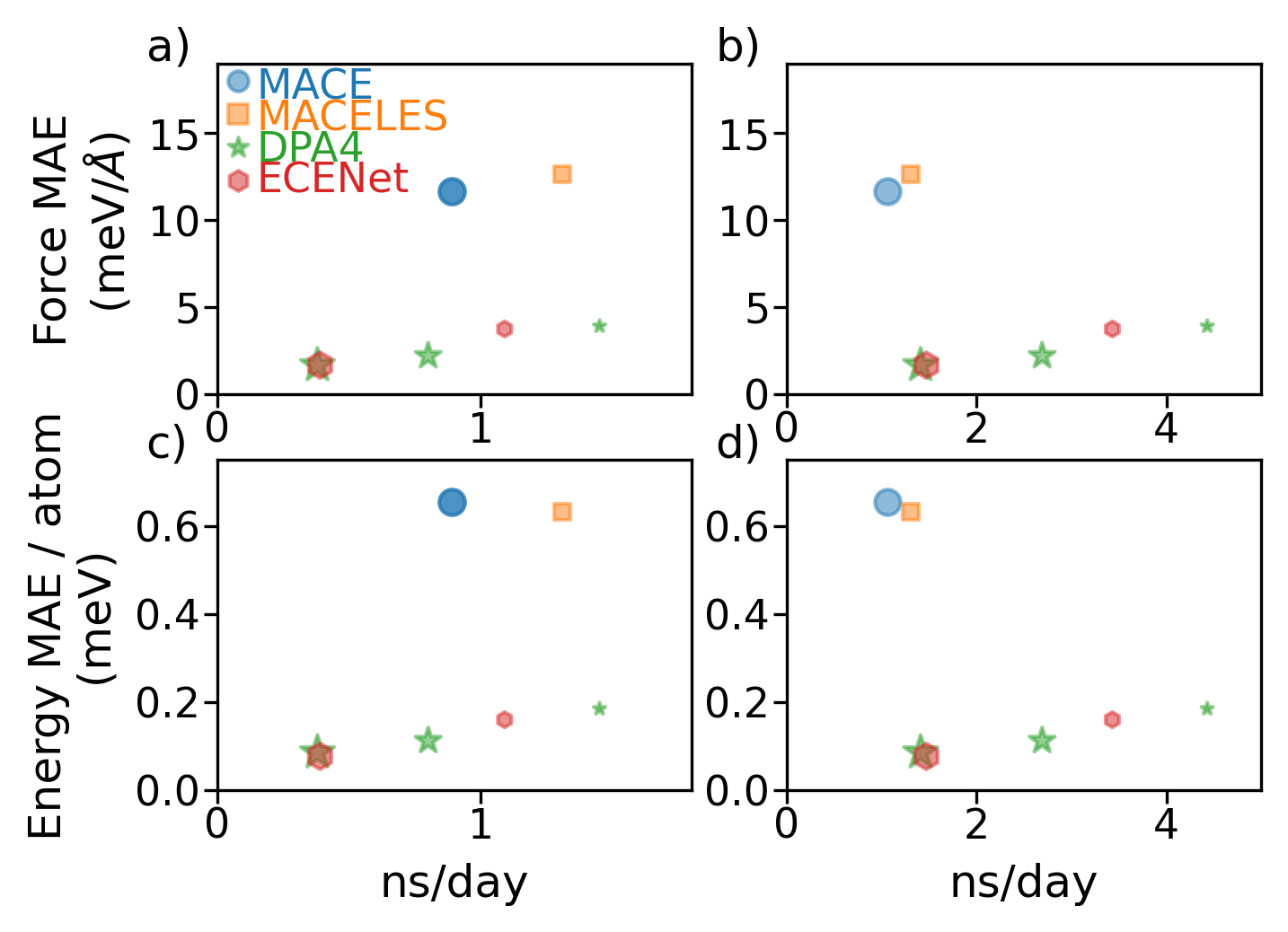}
\caption{\textbf{Pareto accuracy--efficiency frontier for MLIPs trained on the SPICE-MACE-OFF dataset.} We show the geometric mean of the per-split force MAE on the test set against each model's simulation throughput on (a) ice Ih and (b) liquid methanol. 
We do the same for the energy MAE per atom, again with (c) ice-Ih and (d) liquid methanol.
The size of each marker is proportional to the number of parameters in the model; in general, accuracy increases but simulation efficiency decreases as the models grow larger.}
\label{fig:pareto}
\end{figure}

However, comparing test-set accuracy alone can be misleading because more accurate models are generally slower.
For MD, accuracy at a given simulation throughput is more relevant than accuracy alone, since throughput
determines which model can affordably be run in production.
In Figure~\ref{fig:pareto} we therefore plot the geometric-mean force and energy
MAEs from Figure~\ref{fig:spice_results} against the throughput of each model
in NVE simulations of two model systems, ice~I$_h$ and liquid methanol, measured on the same GPU
with the same integrator and time step (see Methods). eSEN and AllScAIP are
absent because their SPICE-trained checkpoints have not been released; the
remaining models were timed with their publicly available weights and each
code's recommended inference settings for efficiency, including the use of TF32 matrix multiplication.

We find that both ECENet models and all three DPA4
models are simultaneously faster and several times more accurate than MACE-4.7M and MACELES-1.9M, and thus the accuracy--efficiency
frontier is defined jointly by ECENet and DPA4. ECENet-4.8M and DPA4-8.8M
occupy essentially the same point, matching each other in both throughput and
error, while ECENet-1.7M sits between DPA4-1.1M and DPA4-5.1M in both speed
and accuracy.
ECENet reaches this frontier with a small set of custom Triton\cite{tillet2019triton} kernels for its edge-frame rotations and nonlinearities, with the remainder of the network running as unfused PyTorch\cite{paszke2019pytorch} operations. Timings and accuracy values in tabular form are provided in Supplementary Table S3.

\subsection{Dipole Prediction and IR Spectrum}
\vspace{-2mm}

\begin{figure}[H]
\centering
\includegraphics[width=0.85\textwidth]{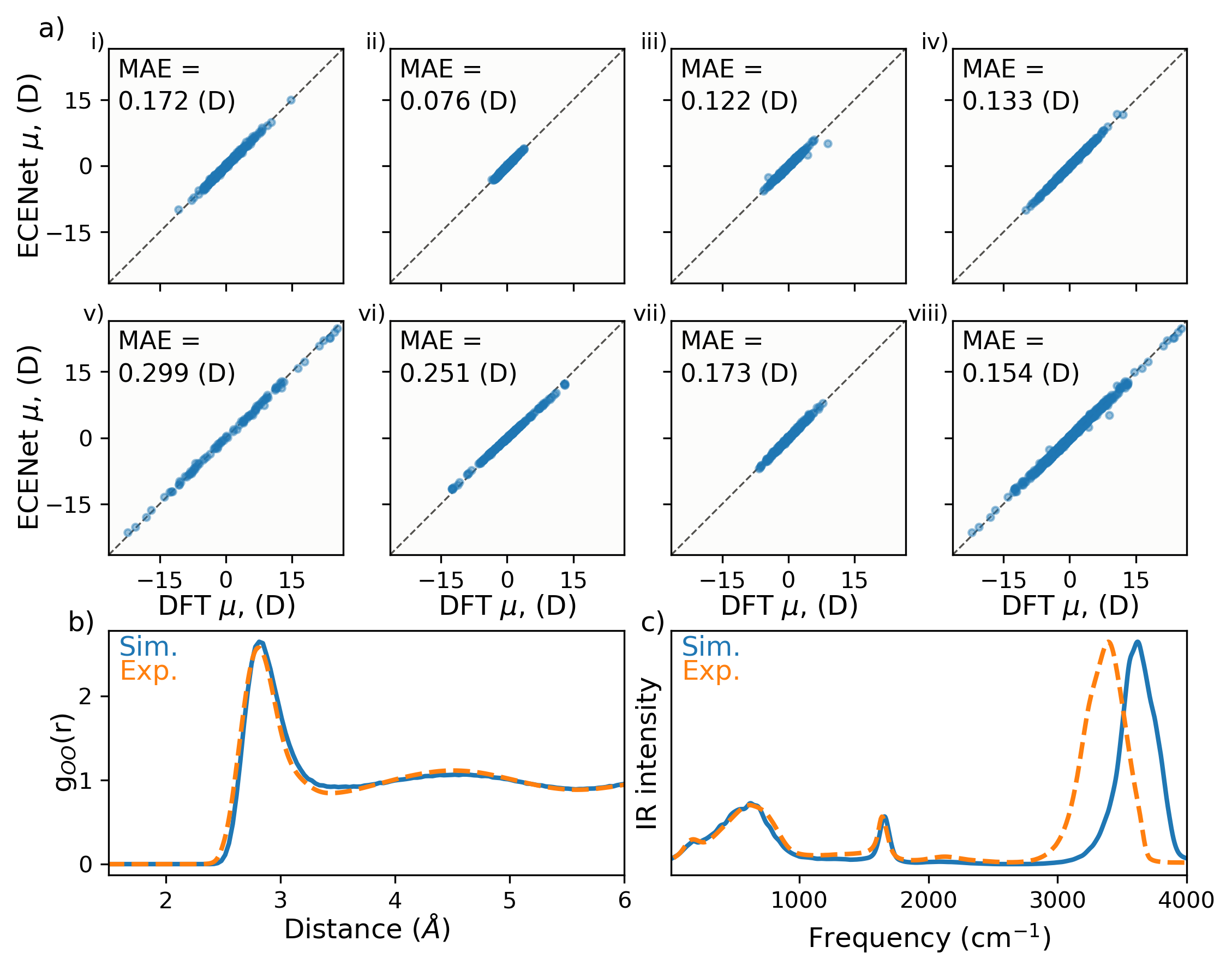}
\vspace{-2mm}
\caption{\textbf{Dipole prediction with ECENet-4.8M}. (a) The dipole moments predicted by ECENet versus the dipole moments computed from DFT for subsets of the SPICE test set. We use the same configurations and dipole moments as Kim et
al.~\cite{kim2025universal} and show each predicted component separately. The panels correspond to (i) PubChem, (ii) DES370K monomers, (iii) DES370K dimers, (iv) dipeptides, (v) solvated amino acids, (vi) water clusters, (vii) QMugs, and (viii) all subsets. (b) The radial distribution function between oxygen atoms ($g_{OO}(r)$) computed from a simulation of liquid water using ECENet-4.8M compared with experiment\cite{Skinner_2014_structure}.
(c) The IR spectrum computed from the same simulation compared with experiment\cite{bertie1996infrared}; see Methods for full calculation details.}
\label{fig:dipoles}
\end{figure}

Incorporating LES\cite{cheng2025latent} gives ECENet latent atomic charges and bond dipoles, and hence molecular dipole moments, even though the model is trained only on energies and forces and never on charges or dipoles. These charges are not guaranteed to be physical: the latent charges are free to take any values that lower the energy and force loss, and they will resemble the true charge
distribution only if that is the most effective way to reproduce the long-range interactions~\cite{cheng2025latent,kim2025universal}. Of the models in Figure~\ref{fig:spice_results}, only MACELES-1.9M, which also uses LES,
produces charges at all. Figure~\ref{fig:dipoles}a compares the molecular dipoles predicted by
ECENet-4.8M with DFT reference values for the SPICE test subsets of Kim et
al.~\cite{kim2025universal}, grouped by category.

Figure~\ref{fig:dipoles}a shows close agreement between the dipole moments
predicted by ECENet and those from DFT across all seven categories,
demonstrating that the latent charges learned from energies and forces alone
are physically meaningful. We next tested whether these charge distributions are meaningful in the condensed phase, as access to dipole moments opens up applications like infrared spectroscopy. We ran NVT molecular dynamics of liquid water at 300~K and the experimental
density with the same SPICE-trained ECENet-4.8M; we emphasize that
the training set contains water only as small gas-phase clusters and no bulk
liquid of any kind. The oxygen–oxygen radial distribution function $g_{OO}(r)$ (Figure 5b) agrees closely with the X-ray diffraction reference of Skinner et al.\cite{Skinner_2014_structure} and lies within the family of radial distribution functions consistent with experimental uncertainty, as analyzed by Brookes and Head-Gordon\cite{Brookes_2015_family}.
Following Kim et al.~\cite{kim2025universal}, we obtained the infrared spectrum from the
autocorrelation of the polarization current,
\begin{equation}
    I(\omega) \propto \int_{0}^{\infty}
    \big\langle \mathbf{J}(0)\cdot\mathbf{J}(t)\big\rangle
    \cos(\omega t)\,\mathrm{d}t,
    \qquad
    \mathbf{J}(t) = \sum_{i=1}^{N} \mathbf{Z}^{*}_{i}(t)\,\mathbf{v}_{i}(t),
    \label{eq:ir}
\end{equation}
where $\mathbf{v}_i$ is the velocity of atom $i$ and $\mathbf{Z}^{*}_i =
\partial \mathbf{M}/\partial \mathbf{r}_i$ is its Born effective charge
tensor, obtained by automatic differentiation of the latent polarization
$\mathbf{M}$ and therefore also predicted without any training on charges.
Figure~\ref{fig:dipoles}c compares the resulting spectrum, normalized to the
OH-stretch maximum, with the measured absorption spectrum of liquid water at
25\,$^\circ$C~\cite{bertie1996infrared}. The positions and relative intensities of the libration, bend, and stretch bands are all reproduced.
As expected for classical MD, the OH-stretch band is blueshifted relative to experiment; the peak redshifts when nuclear quantum effects are included~\cite{medders2015infrared}.

\section{Conclusion}

We introduced ECENet, an equivariant line-graph neural network in which the edges of the atomic graph, rather than the atoms, carry the persistent features. ECENet builds on the edge cluster expansion, with each feature living in the
edge-aligned frame, where it transforms under O(2) rather than O(3), so that Clebsch--Gordan couplings can be replaced by cheaper O(2)-equivariant operations. The features are constructed efficiently by forming an
O(3)-equivariant atom-centered basis once per atom and rotating it into each edge's frame with Wigner $D$-matrices, and messages are passed equivariantly
between edges through their shared atoms, again using Wigner $D$-matrices to move features into and out of edge frames. This equivariant line-graph design differs from existing MLIPs and yields a model whose
number of features scales with the number of edges while its parameter count
does not.

We examined ECENet on several paradigmatic datasets and consistently obtained
results at or near the state of the art. Training on individual systems showed
that equivariant edge-to-edge message passing and LES
are complementary, with systems with strong electrostatic interactions (e.g., water) benefiting more from LES and dispersion-dominated systems (e.g., organic molecules) benefiting more from message-passing layers.
On the MD22 dataset of biomolecules and supramolecular complexes, ECENet matches or exceeds the previous top performer on most energy and force metrics with a fraction of the parameters. On the SPICE-MACE-OFF dataset,
ECENet lies on the accuracy--efficiency frontier alongside DPA4, the strongest current model, matching its accuracy and cost with roughly half the parameters.
The parameter efficiency likely stems from the greater number of features when using an edge-based model as opposed to a node-based one. Because ECENet handles long-range electrostatics through latent Ewald summation, it also predicts quantities that a purely short-range model cannot. Molecular dipoles on the SPICE test set agree closely with DFT across all chemical categories, and the same model, trained only on isolated molecules and small clusters, reproduces the structure and the infrared spectrum of liquid water, demonstrating that charges and dipoles inferred from energies and forces alone are accurate enough to be used, not merely to improve the fit. 

The ECENet-SPICE models are accurate and efficient potentials for organic systems in their current form, but we expect accuracy to improve further with dataset
size and are thus scaling ECENet to larger molecular and materials datasets. Those experiments will test whether an edge-based, O(2)-equivariant representation remains parameter-efficient as chemical diversity grows; our results on SPICE-MACE-OFF suggest
that it will.

\section*{Methods}
\subsection*{ECENet Model Details}
\paragraph{Radial basis and envelope function.}

We used a radial basis in three places in ECENet: during calculation of the initial atomic basis, as an input for the FiLM injection, and as the vector contracted with the output of the final MLP.
In all cases we used the same basis, a series of sinc functions up to $n_{max}$ (16 in all cases) multiplied by a cosine cutoff envelope:
\begin{equation}
f_{\mathrm{cut}}(r) =
\begin{cases}
\tfrac{1}{2}\left[1+\cos\!\left(\dfrac{\pi r}{r_c}\right)\right], & r < r_c,\\[6pt]
0, & r \ge r_c.
\end{cases}
\end{equation}
We used the same envelope function when aggregating messages in equation \ref{eq:aggregate}. 

\paragraph{Edge frames and Wigner $D$-matrices.}
Real spherical harmonics are evaluated with \texttt{sphericart}\cite{bigi2023sphericart} in the
orthonormal, $z$-polar convention. For each directed edge we construct a right-handed orthonormal frame $\{\mathbf e_x,\mathbf e_y,\mathbf e_z\}$ with $\mathbf e_z=\hat{\mathbf r}_{ij}$ by Gram--Schmidt orthogonalization of a fixed reference axis against $\hat{\mathbf r}_{ij}$; the reference is
$\hat{\mathbf x}$, or $\hat{\mathbf y}$ when $|\hat{\mathbf r}_{ij}\cdot \hat{\mathbf x}|\ge 0.9$, so that the construction is free of singularities and large gradients everywhere on the unit sphere. The choice of reference
axis fixes only the azimuthal origin of the frame, a gauge under which every subsequent operation is O(2)-equivariant and the readout (which uses the m = 0 channels) is invariant, so predictions do not depend on it and remain
continuous where the two charts meet. The $\ell=1$ Wigner matrix $D^{1}(\hat{\mathbf r}_{ij})$ is this frame's rotation matrix expressed in the real spherical-harmonic basis, obtained directly from the components of
$\hat{\mathbf r}_{ij}$ without Euler angles or trigonometric functions. Matrices for $\ell\ge 2$ follow from the Clebsch--Gordan recursion\cite{ivanic1996rotation, ivanic1998rotation}:
\begin{equation}
  D^{\ell} = C_{1,\ell-1}^{\ell\,\top}\,\big(D^{1}\otimes D^{\ell-1}\big)\,
  C_{1,\ell-1}^{\ell},
\end{equation}
where $C_{1,\ell-1}^{\ell}$ are the real Clebsch--Gordan coefficients coupling degrees $1$ and $\ell-1$ to $\ell$, precomputed once. The block-diagonal matrix $\bigoplus_{\ell=0}^{\ell_{\max}}D^{\ell}$ is built
once per edge per forward pass and reused for every rotation into the edge frame and, through its transpose, for the rotation of messages back to the global frame. Because the frame is a differentiable function of the atomic positions, forces obtained by automatic differentiation include the dependence of the rotation on the edge direction.

\paragraph{Long-range electrostatics.}
The long-range energy is evaluated with the reference implementation of latent Ewald summation~\cite{cheng2025latent,kim2025universal} (LES). Latent charges and bond dipoles are multiplied by
a fixed factor of $0.1$ before entering the long-range energy so that it
starts small relative to the short-range fit. Charges are smeared with
Gaussians of width $\sigma=1.5$~\AA. For periodic systems the charge--charge,
charge--dipole and dipole--dipole interactions are evaluated by Ewald
summation with a real-space grid spacing of $2.0$ \AA{} and the standard neutralizing background;
for isolated molecules they are evaluated directly in real space.

\begin{table}[h]
\centering
\caption{Model hyperparameters. $c$ is the number of channels per $(\ell,m)$
in the atomic basis, $c'$ the bottleneck width of the
equivariant blocks, $N$ the azimuthal grid size, $H$ the number of gating
heads, and $c_{\mathrm{MP}}$ the width of the message and receiving blocks.}
\label{tab:hparams}
\begin{tabular}{lcccc}
\hline
 & Figure \ref{fig:range} systems & MD22 & ECENet-1.7M & ECENet-4.8M \\
\hline
cutoff $r_c$ (\AA)        & 3--6    & 6      & 5      & 5 \\
$\ell_{\max}$             & 2       &  2  & 3 & 3 \\
$m_{\max}$                & 2       & 2      & 2      & 2 \\
$n_{\max}$                & 16      & 16 & 16     & 16 \\
$c$                       & 32      & 32  & 24 & 42 \\
$c'$                      & 128      & 128  & 256    & 256 \\
$N$ (grid points)         & 10      & 10 & 10     & 16 \\
$N_{\mathrm{MP}}$         & 0--3    & 2      & 1      & 3 \\
$c_{\mathrm{MP}}$, $H$    & 64, 8   & 64, 8 & 64, 8  & 128, 6 \\
energy MLP         & [256, 256] & [256, 256] & [512, 512] & [512, 512] \\
FiLM $d_{\mathrm{el}}$, MLP & 4, [64, 128] & 4, [64, 128] & 10, [128, 128, 128] & 10, [128, 128, 128] \\
LES                       & with/without & yes & yes & yes \\
parameters                & 0.4--1.3M & 1.1M & 1.7M & 4.8M \\
\hline
\multicolumn{5}{l}{\textit{Training}} \\
loss                          & $\mathcal H_{0.0025}$    & $\mathcal H_{0.0025}$  & $\mathcal H_{0.0025}$ & $\mathcal H_{0.0025}$ \\
initial $w_E$, $w_F$                  & 0.1, 1 & 0.1, 1 & 0.1, 1 & 0.1, 1 \\
fine-tune $w_E$, $w_F$                  & NA   & 20, 0.5 & 10, 0.5 & 10, 0.5 \\
optimizer                     & AdamW      & AdamW   & AdamW  & AdamW \\
initial learning rate         & $10^{-3}$  & $10^{-3}$ & $5\times10^{-4}$ & $5\times10^{-4}$ \\
weight decay                  & $10^{-8}$  & $10^{-8}$  & $10^{-8}$ & $10^{-8}$ \\
gradient clipping             & 1.0  & 1.0   & 1.0    & 1.0 \\
warm-up epochs                & 0  & 0  & 1      & 1 \\
epochs       & 800 & 800 + 200 & 200 + 20 & 300 + 20 \\
batch                         & 4 structures & 2--4 structures & 250 atoms/rank & 250 atoms/rank \\
GPUs                          & 1 A100 & 1 A100  & 16 A100 & 16 A100 \\
\hline
\end{tabular}
\end{table}

\subsection*{Training}

All models were trained on energies and forces with the AdamW optimizer
($\beta=(0.9,0.999)$, weight decay $10^{-8}$) on a loss
$\mathcal{L}=w_E\,\mathcal{H}_\delta\!\big((E_{\mathrm{pred}}-E_{\mathrm{ref}})/N_{\mathrm{at}}\big)
+w_F\,\mathcal{H}_\delta\!\big(\mathbf F_{\mathrm{pred}}-\mathbf F_{\mathrm{ref}}\big)$,
where the energy error is taken per atom and averaged over structures, the
force error is averaged over Cartesian components, and $\mathcal H_\delta$
is the Huber loss (quadratic below $\delta$, linear above). 
The units of $\mathcal H_{\delta}$ are eV/atom and
eV/\AA{} for the SPICE-trained models and kcal/mol/atom
and kcal/mol/\AA{} for the others.
Forces are
obtained by automatic differentiation of the total energy, including the
long-range term, so that the short-range network and the latent charges were trained jointly on one computational graph. 
We used a milestone-based learning rate schedule, in which the learning rate  is halved at specific epochs.

\paragraph{Range exploration systems (Figure \ref{fig:range}).} 
For liquid water, we used 574 training, 30 validation, and 50 test configurations taken from reference \citenum{schmiedmayer2024derivative}. 
For Au$_2$ on MgO we used 4250 training, 250 validation, and 500 test configurations taken from
reference ~\citenum{ko2021fourth}. 
Models were trained for 800 epochs in single precision with a batch size of 4. 
Learning rate milestones were placed at epochs 400, 600, 700, 750, 775, and 790.

\paragraph{MD22.}
We used the training set sizes of reference \citenum{chmiela2023accurate}, which the baselines in Table \ref{tab:md22} share: 6000 (Ac-Ala3-NHMe), 8000 (DHA), 8000
(stachyose), 3000 (AT-AT), 2000 (AT-AT-CG-CG), 600 (buckyball catcher) and 800 (double-walled nanotube) configurations, with 5$\%$ held out for validation in each case. We used 200 frames from the same trajectory for the test set. 
Models were trained in single precision for 800 epochs and then fine-tuned with a higher energy weight for a further 200 epochs.
Stachyose, the buckyball catcher and the double-walled nanotube were trained with a batch size of 2; other models were trained with a batch size of 4.
Learning rate milestones were placed at epochs 400, 600, 700, 750, 775, 790, 850, 900, 950, and 975.

\paragraph{SPICE-MACE-OFF.}
We trained on the SPICE-MACE-OFF split of SPICE~v1~\cite{Eastman_2023_spice, kovacs2025}, holding out 51,005 structures for validation, and report errors on the standard per-category test set. 
Both models were trained using distributed data parallelism on 16 A100 GPUs with size-aware batching to an atom budget of 250 atoms per rank. Models were trained in single precision with TF32 matrix multiplication for 200
epochs (ECENet-1.7M) or 300 epochs (ECENet-4.8M), then fine-tuned for a further 20 epochs in full single precision (TF32 disabled). For ECENet-1.7M, learning rate milestones were placed at 80, 130, 170, 190, 210, and 215. For ECENet-4.8M, learning rate milestones were placed at 128, 192, 256, 280, 290, 310, and 315. Per-category MAEs are reported for the forces and the energies per atom; the geometric mean over the seven categories is used as the single-number summary.

\subsection*{Throughput benchmarks}

All timings in Figure \ref{fig:pareto} were obtained on a single NVIDIA A100
40~GB GPU, running NVE molecular
dynamics in the Atomic Simulation Environment\cite{Larsen_2017_ase} with the velocity-Verlet integrator at a 1.0~fs time step. 
Each measurement comprised 10 warm-up steps followed by 100 timed steps, and
throughput is reported as ns/day. Every model was run in float32 with TF32
enabled, from its publicly released weights.
ECENet used its fused Triton kernels for the edge-frame rotations and the real-space nonlinearity; DPA4 was frozen with \texttt{dp freeze} on the target GPU; and MACE and MACELES used the MACE calculator with cuEquivariance acceleration. The SPICE-trained eSEN and AllScAIP checkpoints have not been released, so those models appear in Figure \ref{fig:spice_results} but not in Figure \ref{fig:pareto}. The two condensed-phase systems fall under the 100-neighbor limit of the DPA4-SPICE models.

\paragraph{Liquid water simulation and IR spectrum (Figure 5b,c).}
We simulated 80 water molecules in a cubic cell of side 13.389~\AA{}
(0.997~g/cm$^{3}$) with ECENet-4.8M, using the Atomic Simulation Environment\cite{Larsen_2017_ase} Langevin integrator
at 300~K with a friction coefficient of $10^{-3}$~fs$^{-1}$ and a
0.5~fs time step, starting from Maxwell--Boltzmann velocities at 300~K.
The model was evaluated in single precision with TF32 matrix
multiplication. The trajectory was run for 275 ps, with the first 175 ps discarded as equilibration.
$g_{\mathrm{OO}}(r)$ was computed from the remaining frames. The
polarization current $\mathbf{J}(t)$ of equation \ref{eq:ir} was evaluated from the
per-frame Born effective charges and velocities, its autocorrelation
function was smoothed with a Gaussian window, and the resulting spectrum was normalized to the
maximum of the OH-stretch band.

\section*{Data Availability}
\vspace{-2mm}

\noindent The Au$_2$ on MgO dataset is taken from reference  \citenum{ko2021fourth}; the water dataset from reference \citenum{schmiedmayer2024derivative}; the MD22 dataset from reference \citenum{chmiela2023accurate}; and the SPICE-MACE-OFF dataset from reference\citenum{kovacs2025}.
Tabular data for the models' performance on the SPICE-MACE-OFF dataset are given in the supplementary information.
Model weights for the SPICE-trained ECENet models can be found at \texttt{https://huggingface.co/THGLab/ECENet-4.8M-SPICE} and \newline \texttt{https://huggingface.co/THGLab/ECENet-1.7M-SPICE}.

\section*{Code Availability}
\vspace{-2mm}

\noindent All ECE architecture code can be found at \texttt{https://github.com/THGLab/ecenet}. 

\section*{Acknowledgments}
\vspace{-2mm}

\noindent We thank Dongjin Kim and Bingqing Cheng for helpful discussions in regards to LES. We thank the CPIMS program, Office of Science, Office of Basic Energy Sciences, Chemical Sciences Division of the U.S. Department of Energy under Contract DE-AC02-05CH11231 for support of the machine learning.  This research used computational resources of the National Energy Research Scientific Computing, a DOE Office of Science User Facility supported by the Office of Science of the U.S. Department of Energy under Contract No. DE-AC02-05CH11231.

Anthropic Claude assisted with literature organization,
manuscript editing, and code development. The authors directed its use and independently verified all scientific claims, calculations, code, figures, and references.

\section*{Author Contributions Statement}
\vspace{-2mm}

\noindent R.A.L. and T.H.G. defined goals of the project. R.A.L. created and developed the ECE model and performed all calculations. R.A.L. and T.H.G. discussed the results and wrote and edited the manuscript.

\section*{Competing Interests Statement}
\vspace{-2mm}

\noindent The authors declare no competing interests.

\bibliographystyle{achemso}

\bibliography{references}

\end{document}


\title{Supplementary Information\\
ECENet: An Edge Cluster Expansion Line Graph Neural Network}
\date{}
\author{R. Allen LaCour*$^{1,2}$, Teresa Head-Gordon$^{*1,2,3}$}
\maketitle
\noindent
\begin{center}
$^1$Kenneth S. Pitzer Theory Center and Department of Chemistry\\
$^2$Chemical Sciences Division, Lawrence Berkeley National Laboratory\\
$^3$Departments of Bioengineering and Chemical and Biomolecular Engineering\\
University of California, Berkeley, CA, 94720 USA

corresponding author: alacour@berkeley.edu, thg@berkeley.edu
\end{center}

\section*{Supplementary Tables}
\begin{table}[h]
\centering
\footnotesize
\setlength{\tabcolsep}{3.5pt}
\caption{Per-category force MAE on the MACE-OFF SPICE test set, in meV/\AA{} (the data of Figure 3). The last column is the geometric mean over the seven categories (LWAMAE). Best value per column in bold. Superscripts give the source of the reported values; rows without a superscript are this work.}
\label{tab:spice_forces}
\begin{tabular}{lcccccccc}
\hline
Model & PubChem & \shortstack{DES370K\\monomers} & \shortstack{DES370K\\dimers} & Dipeptides & \shortstack{Solvated\\amino acids} & Water & QMugs & \shortstack{Geometric\\mean} \\
\hline
MACE-4.7M\textsuperscript{\citenum{kovacs2025}} & 14.75 & 6.58 & 6.62 & 10.19 & 19.43 & 16.93 & 13.57 & 11.66 \\
MACELES-1.9M\textsuperscript{\citenum{kim2025universal}} & 17.24 & 7.65 & 7.48 & 11.46 & 18.37 & 12.54 & 20.11 & 12.67 \\
eSEN-3.2M\textsuperscript{\citenum{fu2025learning}} & 6.10 & 1.85 & 2.77 & 3.04 & 5.76 & 3.88 & 5.70 & 3.83 \\
eSEN-6.5M\textsuperscript{\citenum{fu2025learning}} & 4.21 & 1.24 & 2.12 & 2.00 & 3.68 & 2.50 & 3.78 & 2.58 \\
DPA4-1.1M\textsuperscript{\citenum{li2026dpa4}} & 5.92 & 2.13 & 2.16 & 3.10 & 6.69 & 4.46 & 5.36 & 3.89 \\
DPA4-5.1M\textsuperscript{\citenum{li2026dpa4}} & 3.59 & 1.22 & 1.23 & 1.68 & 3.59 & 2.47 & 2.91 & 2.18 \\
DPA4-8.8M\textsuperscript{\citenum{li2026dpa4}} & \textbf{2.98} & 0.91 & 0.96 & \textbf{1.25} & 2.59 & 1.78 & \textbf{2.14} & \textbf{1.64} \\
AllScAIP-34M\textsuperscript{\citenum{qu2026recipe}} & 3.21 & 0.97 & 0.95 & 1.54 & 4.86 & 2.12 & 2.73 & 2.00 \\
ECENet-1.7M & 6.88 & 1.88 & 1.73 & 3.53 & 5.88 & 3.41 & 6.35 & 3.73 \\
ECENet-4.8M & 3.42 & \textbf{0.88} & \textbf{0.79} & 1.52 & \textbf{2.37} & \textbf{1.45} & 2.66 & 1.65 \\
\hline
\end{tabular}
\end{table}

\begin{table}[h]
\centering
\footnotesize
\setlength{\tabcolsep}{3.5pt}
\caption{Per-category energy MAE on the MACE-OFF SPICE test set, in meV/atom (the data of Figure 3). The last column is the geometric mean over the seven categories (LWAMAE), given to three decimals. Best value per column in bold. Superscripts give the source of the reported values; rows without a superscript are this work.}
\label{tab:spice_energies}
\begin{tabular}{lcccccccc}
\hline
Model & PubChem & \shortstack{DES370K\\monomers} & \shortstack{DES370K\\dimers} & Dipeptides & \shortstack{Solvated\\amino acids} & Water & QMugs & \shortstack{Geometric\\mean} \\
\hline
MACE-4.7M\cite{kovacs2025} & 0.88 & 0.59 & 0.54 & 0.42 & 0.98 & 0.83 & 0.54 & 0.655 \\
MACELES-1.9M\cite{kim2025universal} & 0.71 & 0.54 & 0.47 & 0.52 & 0.82 & 0.69 & 0.76 & 0.632 \\
eSEN-3.2M\cite{fu2025learning} & 0.22 & 0.17 & 0.20 & 0.10 & 0.30 & 0.24 & 0.16 & 0.189 \\
eSEN-6.5M\cite{fu2025learning} & 0.15 & 0.13 & 0.15 & 0.07 & 0.25 & 0.15 & 0.12 & 0.137 \\
DPA4-1.1M\cite{li2026dpa4} & 0.24 & 0.17 & 0.18 & 0.10 & 0.27 & 0.30 & 0.12 & 0.184 \\
DPA4-5.1M\cite{li2026dpa4} & 0.13 & 0.17 & 0.11 & 0.06 & 0.13 & 0.16 & 0.07 & 0.111 \\
DPA4-8.8M\cite{li2026dpa4} & 0.12 & 0.12 & 0.11 & \textbf{0.04} & 0.09 & 0.11 & \textbf{0.05} & 0.085 \\
AllScAIP-34M\cite{qu2026recipe} & 0.13 & \textbf{0.07} & \textbf{0.06} & 0.06 & 0.13 & \textbf{0.08} & 0.11 & 0.087 \\
ECENet-1.7M & 0.22 & 0.15 & 0.13 & 0.11 & 0.19 & 0.17 & 0.17 & 0.159 \\
ECENet-4.8M & \textbf{0.11} & 0.10 & \textbf{0.06} & 0.05 & \textbf{0.07} & 0.09 & 0.07 & \textbf{0.076} \\
\hline
\end{tabular}
\end{table}

\begin{table}[h]
\centering
\footnotesize
\setlength{\tabcolsep}{5pt}
\caption{Simulation throughput and accuracy of the models (the data of Figure 4). Throughput is the NVE molecular-dynamics rate in ns/day on a single NVIDIA A100 for ice~I$_h$ (1536 atoms) and liquid methanol (384 atoms), measured under the common protocol described in the Methods (float32 with TF32, DPA4 frozen, MACE with cuEquivariance). Accuracy is the geometric mean over the seven MACE-OFF SPICE test categories of the force MAE (meV/\AA) and energy MAE (meV/atom) from Tables~\ref{tab:spice_forces} and \ref{tab:spice_energies}. Superscripts give the source of the accuracy values; throughputs were all measured in this work. eSEN and AllScAIP are absent because their SPICE-trained checkpoints have not been released.}
\label{tab:pareto}
\begin{tabular}{lcccc}
\hline
Model & \shortstack{Ice I$_h$\\(ns/day)} & \shortstack{Methanol\\(ns/day)} & \shortstack{Force MAE\\geo.\ mean (meV/\AA)} & \shortstack{Energy MAE\\geo.\ mean (meV/atom)} \\
\hline
MACE-4.7M\cite{kovacs2025} & 0.92 & 1.38 & 11.66 & 0.655 \\
MACELES-1.9M\cite{kim2025universal}& 1.31 & 1.86 & 12.67 & 0.632 \\
DPA4-1.1M\cite{li2026dpa4} & 1.45 & 4.43 & 3.89 & 0.184 \\
DPA4-5.1M\cite{li2026dpa4}& 0.80 & 2.69 & 2.18 & 0.111 \\
DPA4-8.8M\textsuperscript{\citenum{li2026dpa4}} & 0.38 & 1.41 & 1.64 & 0.085 \\
ECENet-1.7M & 1.09 & 3.43 & 3.73 & 0.159 \\
ECENet-4.8M & 0.39 & 1.47 & 1.65 & 0.076 \\
\hline
\end{tabular}
\end{table}

\newpage
\bibliographystyle{achemso}
\bibliography{references}